\documentclass[12pt]{iopart}
\usepackage{iopams} 
\usepackage{graphicx}
\usepackage{enumerate}

\begin{document}

\title[Thermal properties of composite materials]{Thermal properties of composite materials with a complex fractal structure}

\author{F  Cervantes-\'Alvarez$^1$, J J Reyes-Salgado$^1$, V Dossetti$^{1, 2}$ and \hbox{J L Carrillo$^1$}}

\address{$^1$ Instituto de F\'{\i}sica, Benem\'erita Universidad Aut\'onoma de Puebla, \\Apartado Postal J-48, Puebla 72570, Mexico}
\address{$^2$ Consortium of the Americas for Interdisciplinary Science and Department of Physics and Astronomy, University of New Mexico, Albuquerque, NM 87131, USA}
\ead{dossetti@ifuap.buap.mx}

\begin{abstract}
In this work, we report the thermal characterization of platelike composite samples made of polyester resin and magnetite inclusions. By means of photoacoustic spectroscopy and thermal relaxation, the thermal diffusivity, conductivity and volumetric heat capacity of the samples were experimentally measured. The volume fraction of inclusions was systematically varied in order to study the changes in the effective thermal conductivity of the composites. In some samples, a static magnetic field was applied during the polymerization process resulting in anisotropic inclusion distributions. Our results show a decrease in the thermal conductivity of some of the anisotropic samples compared to the isotropic randomly distributed ones. Our analysis indicates that the development of elongated inclusion structures leads to the formation of magnetite and resin domains causing this effect. We correlate the complexity of the inclusion structure with the observed thermal response by a multifractal and lacunarity analysis. All the experimental data are contrasted with the well known Maxwell-Garnett's effective media approximation for composite materials.
\end{abstract}


\section{Introduction}
\label{sec:intro}

Tuning the effective thermal properties of composite ma\-te\-rials, by controlling their internal structure and composition, is a topic of great interest nowadays due to potential applications that span many different kinds of systems: from fusion reactors to electronic devices. Experimentally, this has been achieved by varying the relative volume fraction and size of the constituents \cite{geo04, ago08, goy08}, by controlling the microstructure and interface properties of the inclusions \cite{tag05}, by mechanical strains (stretching the material to orient one of the constituents in the stretching direction) \cite{liu10}, by controlling the morphology of the inclusions \cite{vin12, bou13}, and even by engineering multilayered composites where the diffusive heat flow is guided in order to obtain a desired thermal conduction \cite{nar12}, among others.

On the other hand, experimental techniques such as the photoacoustic (PA) technique in combination with the thermal re\-laxa\-tion method (TRM) have proven to be reliable and useful tools to measure the thermal properties such as the thermal diffusivity, conductivity and heat capacities of materials \cite{dos02}. In particular, the photoacoustic technique has been used to characterize the thermal transport properties (thermal diffusivity and effusivity) of materials present in many different forms like transparent liquids, polymers, powdered and solid samples \cite{geo04, ros76, per87, yan95, ber01, geo01, geo02, don02, var02, hai12, mar12}. The principle of  the PA effect is based  on the light-induced heat release and consequent generation of acoustic waves from a material, when it is irradiated with a modulated optical radiation. Complimentarily, the TRM allows the measurement of heat capacities of small samples in the temperature range around room temperature \cite{hat79}. Experimentally, the sample is placed inside a closed cell where a partial vacuum has been established. Then, the sample is subjected to a constant illumination so that its temperature raises, reaching a maximum and thermal equilibrium is established between its illuminated and non-illuminated faces. Afterwards, the illumination is interrupted and the sample is let to cool down mainly through radiative processes in a characteristic time known as the relaxation time. By measuring the difference between the maximum and room temperatures and the relaxation time, it is possible to determine the volumetric heat capacity of the sample \cite{dos02}. From the application of both techniques, the thermal conductivity of different materials can be obtained.

In this work, by means of the PA technique in combination with the TRM, we study the thermal properties of composite platelike samples consisting in a polyester resin matrix with powdered magnetite inclusions. Three kinds of inclusion structures where considered: one isotropic and two anisotropic. For the isotropic samples, the inclusion dispersion was prepared to be as random as possible. For the anisotropic samples, a magnetic field was applied during the polymerization process, resulting in two diffe\-rent kinds of inclusion structures depending on how the chains formed by the magnetite particles are oriented: perpendicular or parallel to the faces of the platelike samples. The \hbox{volume} fraction of inclusions was systematically varied in all cases in order to observe the effect on the effective thermal conductivity of the samples. Our results show that, as the volume fraction of inclusions increases, the thermal conductivity of the isotropic samples behaves according to the well known Maxwell-Garnett's effective medium approxi\-ma\-tion for composite materials with a random distribution of spherical inclusions \cite{cho99}. In contrast, the thermal conductivity of the anisotropic \emph{transversal} samples (those were the magnetite chains run parallel to the faces of the samples but transversal to the illumination direction) is always smaller than the isotropic ones with a non-trivial behavior. For the anisotropic \emph{longitudinal} samples (in these the magnetite chains run perpendicular to the faces of the samples but along the illumination direction), those with lower volume fraction of inclusions behave in the same way as the isotropic ones, however, with further increases of the concentration of inclusions the thermal conductivity drops. Our analysis shows that the formation of large chains of magnetite particles in some of the samples leads to the development of (inclusion and matrix) domains, causing the effective thermal conductivity to decrease in accordance to recent theoretical results \cite{yu06}.

\begin{figure}[t]
\begin{center}
\includegraphics[width=0.5\textwidth]{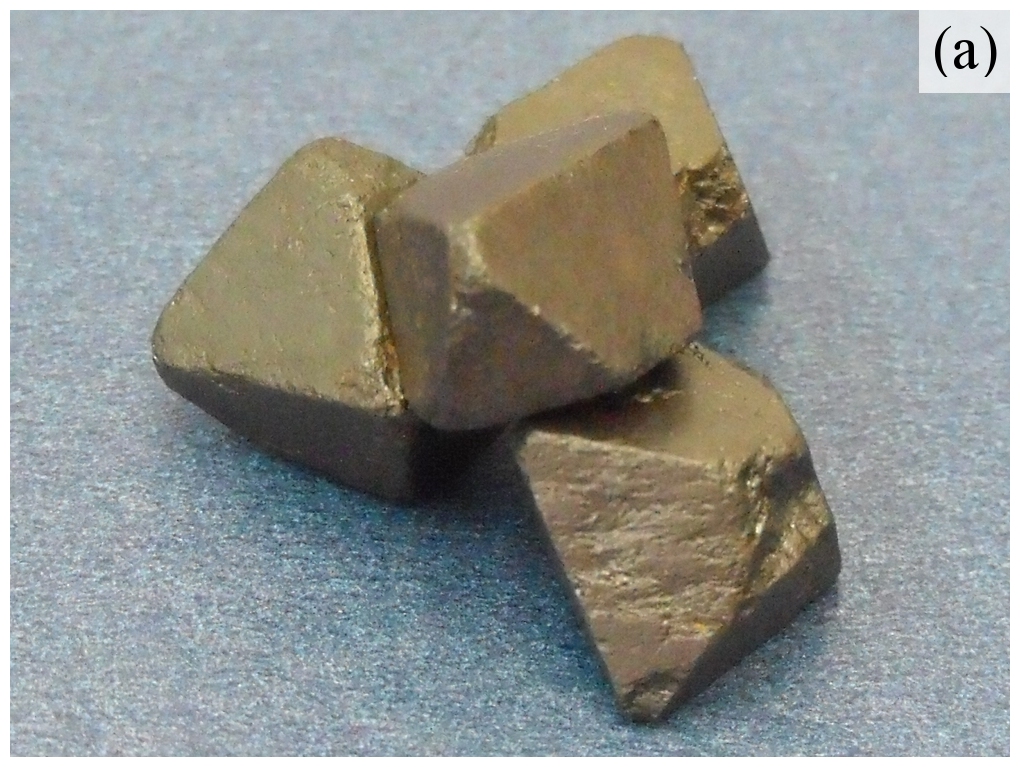}

\includegraphics[width=0.5\textwidth]{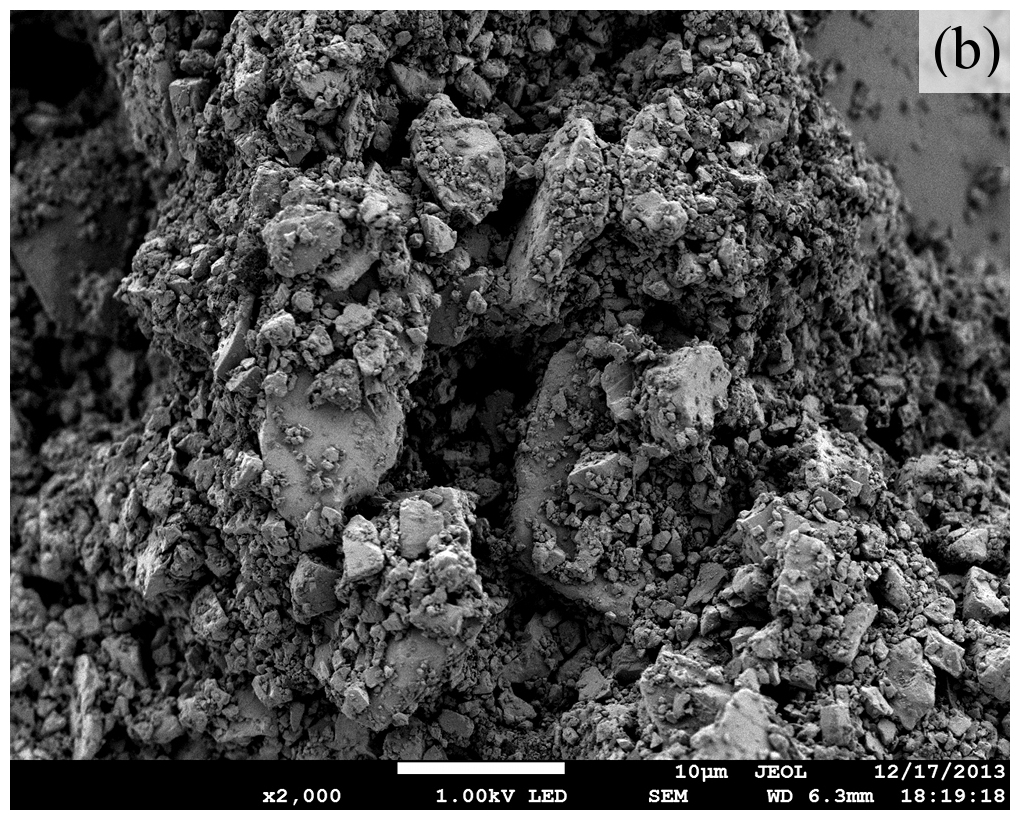}

\includegraphics[width=0.5\textwidth]{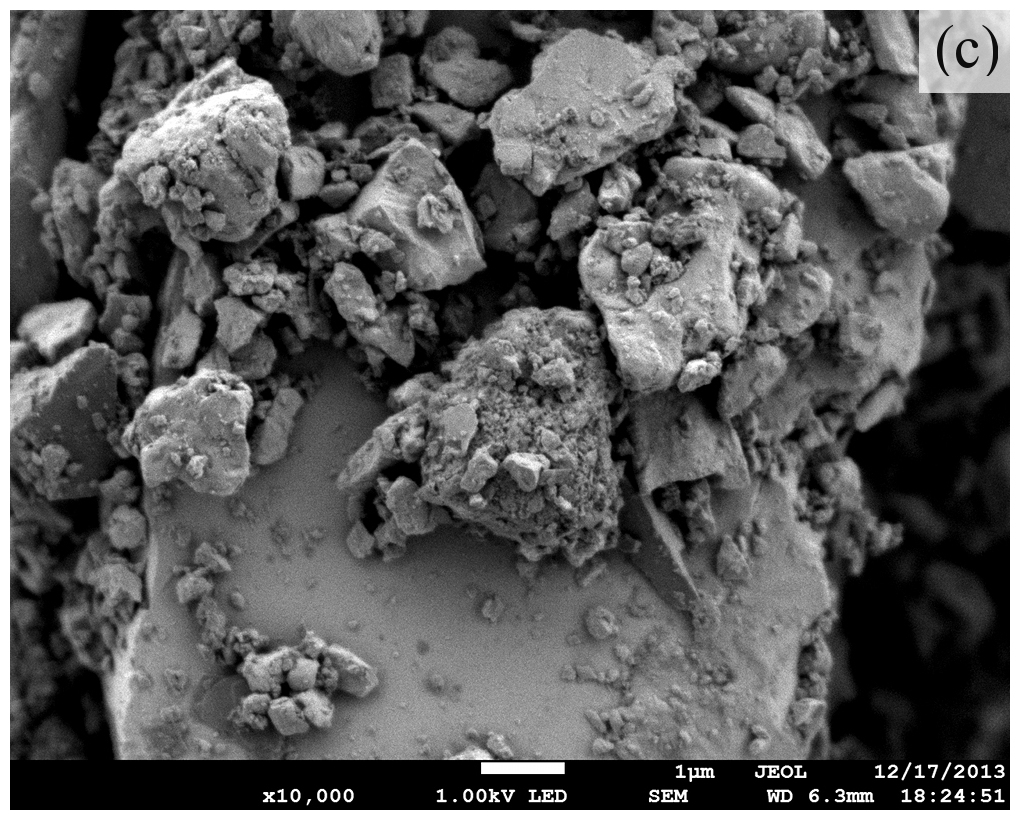}
\caption{In (a), photo of the mineral magnetite crystals from which the inclusions were obtained. In (b) and (c), scanning electron micrographs of the resulting magnetite powder from the grinding and sifting process for two magnifications, X2000 and X10000, respectively. Notice the polydispersity of the grains that goes from nanometric to micrometric scales.}
\label{fig:incl}
\end{center}
\end{figure}

\section{Experimental details}
\label{sec:expdet}

In this section, we provide details regarding the samples preparation and the experimental setups for the photoacoustic technique and thermal relaxation.

\subsection{Inclusions preparation}
\label{ssec:incprep}

The preparation of the composite samples begins with the obtention of the inclusions from mineral magnetite crystals through a grinding process with the use of an agate mortar and pestle. The mineral magnetite itself is a naturally occurring dark iron oxide (Fe$_3$O$_4$) with molecular weight of 231.55 $\mathrm{g}/\mathrm{mol}$ \cite{crc05}, that only presents one crystalline phase \cite{cor03}. We selected magnetite as the inclusion material due to its magnetic response, given the fact that some of the samples were prepared in the presences of magnetic fields in order to obtain a desired inclusion structure (see below). The magnetite crystals shown in figure~\ref{fig:incl}(a), were crushed in the mortar until the size of the particles obtained was less than $44\mu \mathrm{m}$. We made sure of this by sifting the powder through a mesh sieve. In figures \ref{fig:incl}(b) and \ref{fig:incl}(c),  scanning electron microscope (SEM) micrographs from the resulting powder with different magnification factors are presented. As evident from the figures, the magnetite inclusions are very polydisperse with nanometric to micrometric sizes.

\begin{figure}[t]
\begin{center}
\includegraphics[width=0.6\textwidth]{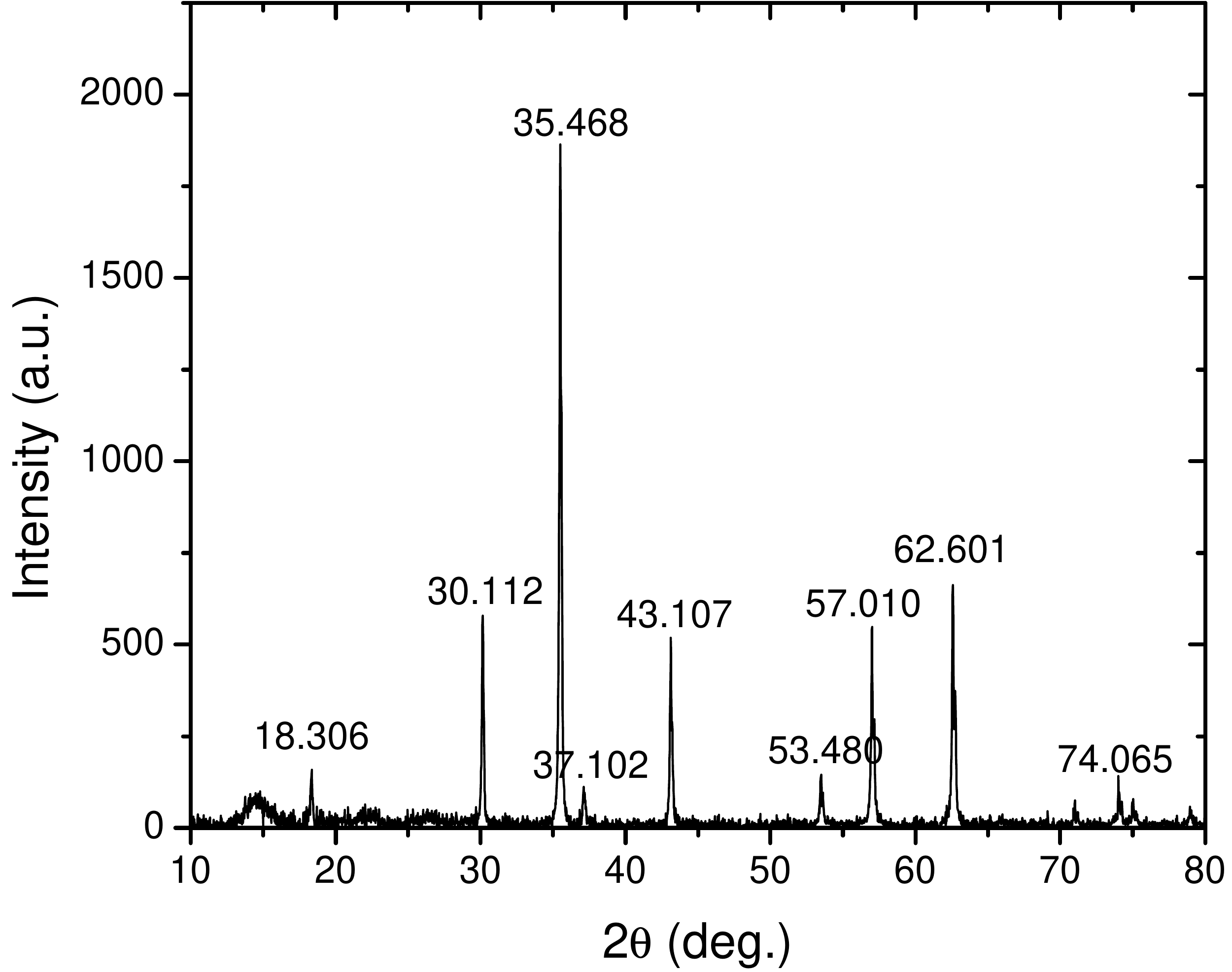} 
\caption{XRD diffractogram obtained for the magnetite powder resulting from the grinding process of the magnetite crystals shown in figure~\ref{fig:incl}(a).}
\label{fig:dif}
\end{center}
\end{figure}

\begin{figure*}
\centering
\includegraphics[width=\textwidth]{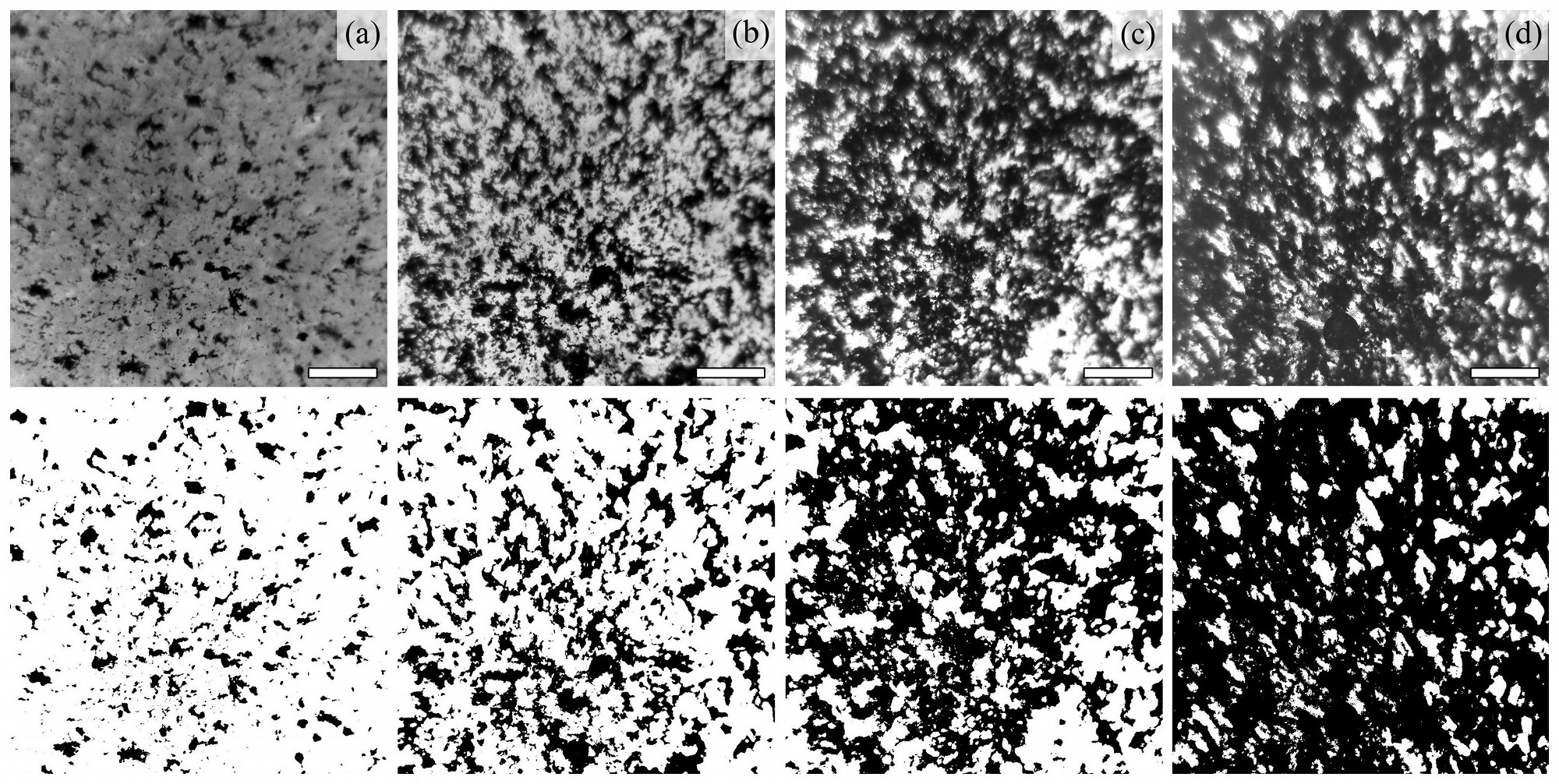}
\includegraphics[width=\textwidth]{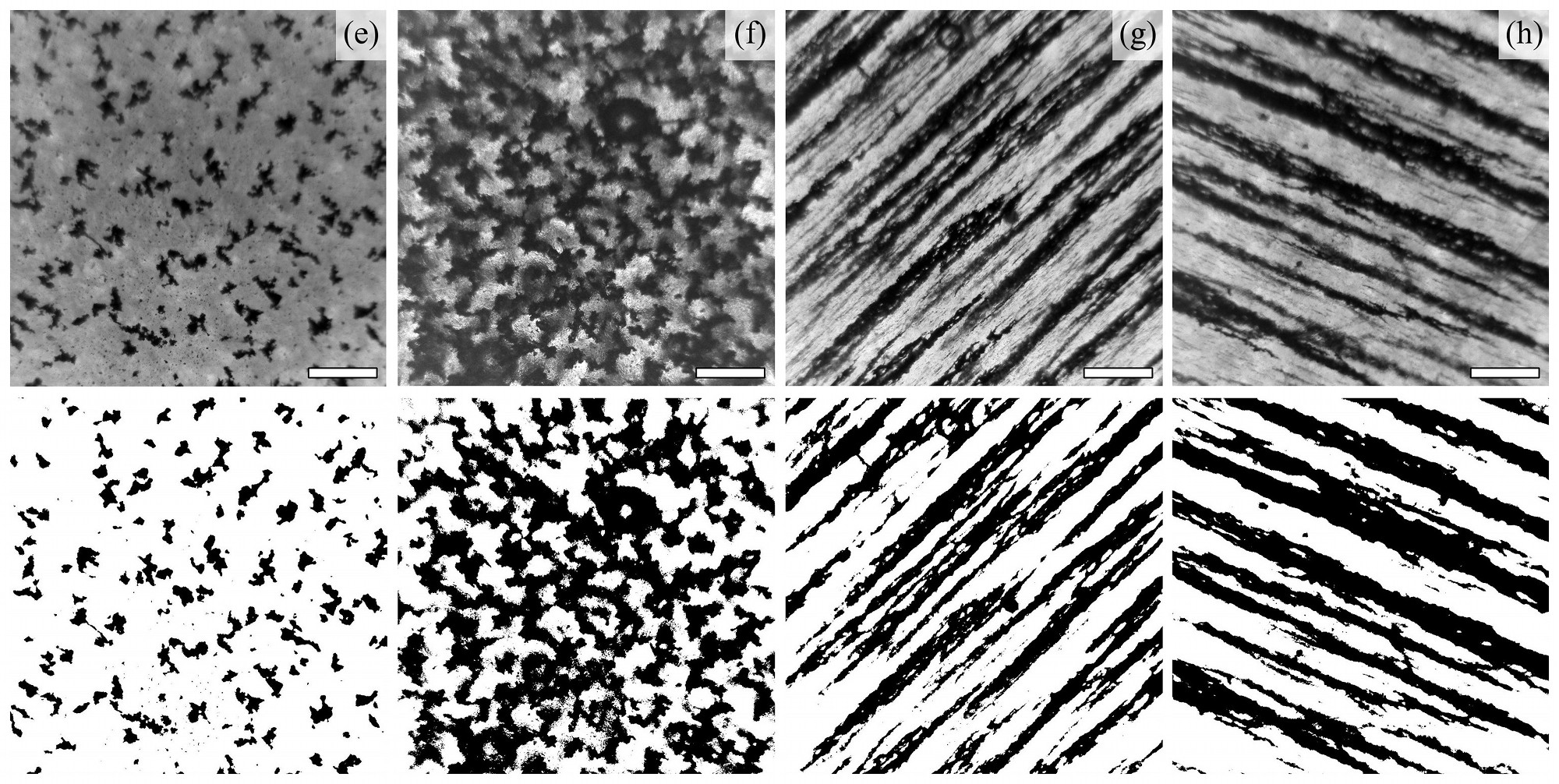}
\caption{Micrographs of composite samples (odd rows) and their corresponding binarized images (even rows), taken with an optical microscope with a magnification X80. The white bar at the bottom right corner of the micrographs corresponds to a scale of 500 $\mu \mathrm{m}$. From (a) to (d), selected isotropic samples with volume fraction concentration of inclusions $\eta_\mathrm{m} = 0.034, 0.066, 0.077, 0.089$, respectively. In (e) and (f), selected anisotropic longitudinal samples with $\eta_\mathrm{m} = 0.042, 0.064$, respectively. In (g) and (h), selected anisotropic transversal samples with $\eta_\mathrm{m} = 0.033, 0.041$, respectively. Notice the formation of large elongated structures of magnetite particles and the development of magnetite and resin domains in (f), (g) and (h). These samples have lower values of thermal conductivity and cannot be described by the Maxwell-Garnett's EMA for composite materials with a random distribution of inclusions (see text for more details).}
\label{fig:ImageJ}
\end{figure*}

Additionally, the magnetite powder was analyzed with an \hbox{X-ray} diffractometer to characterize the magnetite crystals used in this work. Figure~\ref{fig:dif} shows the diffractogram obtained with a PANalytical Empyrean (Cu-k$_\alpha$, $\lambda = 1.5406 \, \mathring{\mathrm{A}}$) diffractometer in $2\theta$ range from $10^{\circ}$ to $80^{\circ}$. By comparing the position and relative amplitude of the narrow peaks in the X-ray diffractogram with the Powder Diffraction File (PDF) with reference code 01-089-0691, we were able to determine that our magnetite powder was made of almost pure magnetite in its crystalline form \cite{zha09}.

\subsection{Samples preparation}
\label{ssec:samprep}

The samples were prepared in bulk in plastic cubic cells of about 3 $\mathrm{cm}^3$ --- one for each kind of inclusion structure and for each concentration of inclusions. The matrix of the samples, consisting in polyester resin, uses a peroxide as catalyzer in order to accelerate its solidification process. The gel time for 100 g of the polyester resin at room temperature is about 14 minutes, while the curing time for the same amount is 22 minutes. Our bulk samples were much smaller than that (about \hbox{3 g}), making the curing time rather fast. Additionally, the viscosity of the resin and the size of our inclusions helped to prevent for any sedimentation process to take place. For these reasons, the magnetite inclusions were aggregated and mixed with the resin before adding the catalyzer so that they could be dispersed as homogeneously as possible inside the matrix. In the case of the anisotropic samples, just after adding the catalyzer, a magnetic field was applied during the polyme\-ri\-za\-tion process through a pair of Helmholtz coils in order to ensure the uniformity of the field. The presence of the magnetic field induced the formation of chains of magnetite particles inside the matrix. The intensity of the magnetic field applied was always 12.17 $\mathrm{kA}/\mathrm{m}$ regardless of the concentration of inclusions.

We let 8 hours pass before taking the bulk samples out of the molds to ensure they have achieved maximal hardness. Then, a section of about 1 $\mathrm{cm}^3$ was obtained from the centre of each bulk sample in order to avoid any possible distortions in the inclusion structure that could come from the surface in contact with the mold itself. Square platelike samples about 1 mm thick and 7 mm in linear size were later obtained from these centre sections. In the case of isotropic samples, a single slice was extracted. In the case of anisotropic samples, two slices were extracted from the centre section: one perpendicular to the direction of the applied field (for the anisotropic \emph{longitudinal} sample) and one parallel to the direction of the applied field (for the anisotropic \emph{transversal} one). Each of the platelike samples was then sanded and polished to the desired thickness with a special die, consisting in a cylindrical body with diameter of about 3.5 cm and a plunger with a diameter of 1 cm, to ensure their faces were parallel. For each kind of composite sample (isotropic, anisotropic longitudinal and anisotropic transversal) the concentration of inclusions $\eta_\mathrm{m}$, measured in volume fraction, was systematically varied within the interval 0.013 to 0.089.

Additionally, two more platelike samples were prepared: one made of resin without inclusions and one made of solid magnetite. These samples were used to determine the thermal properties of the two constituent materials of the composite ones. The results for them were later introduced in the Maxwell-Garnett's effective medium approximation (see below) in order to compare the experimental measurements with the theoretical prediction.

The odd rows in figure~\ref{fig:ImageJ} show micrographs of selected samples, taken with an optical microscope with a magnification of X80, for the three types considered in this work: isotropic in (a) to (d), anisotropic longitudinal in (e) and (f),  and anisotropic transversal in (g) and (h). In order to produce these micrographs, it was necessary to obtain an additional slice from the centre section for each composite sample, this time sanded and polished to a thickness of about $100 \, \mu\mathrm{m}$. This allowed us to observe a few layers of the inclusion structure as can be appreciated in the corresponding binarized images, shown in the even rows of the figure. Afterwards, the inclusion structure was characterized via a multifractal analysis and lacunarity measurements as discussed later in details in section \ref{sec:resdis}.

\begin{figure}[t]
\begin{center}
\includegraphics[width=0.65\textwidth]{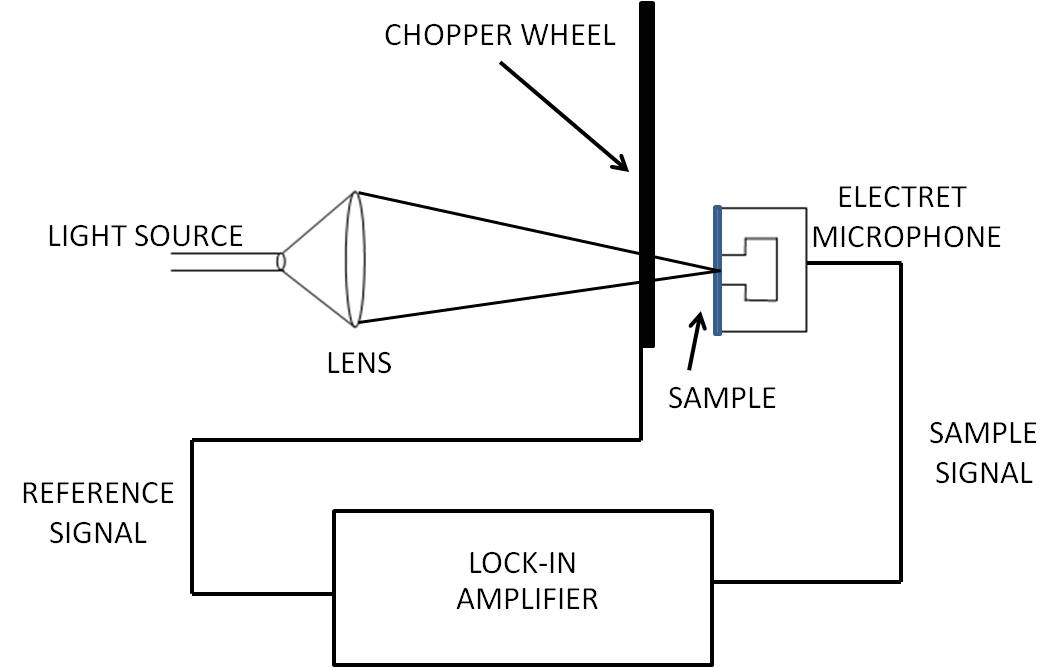} 
\caption{Schematic diagram of the experimental arrangement used to measure the thermal diffusivity, $\alpha_\mathrm{s}$, using the open-cell PA technique (see text for details).}
\label{fig:paset}
\end{center}
\end{figure}

We must mention that the polydispersity of the magnetite particles used in this work induced the aggregation of particles, i.e., large particles tend to get covered by smaller ones; see, for example, figures~\ref{fig:incl}(b) and \ref{fig:incl}(c). This fact prevented a better dissolution of the inclusions in the resin, resulting in less homogenous samples than desired. Because of this and in order to have a more accurate value of the concentration of inclusions, the density of the composite samples was measured and their concentration $\eta_\mathrm{m}$ calculated from $\eta_\mathrm{m} = 1 - (\rho - \rho_\mathrm{m})/(\rho_\mathrm{r} -\rho_\mathrm{m})$, where $\rho$, $\rho_\mathrm{r}$ and $\rho_\mathrm{m}$ correspond to the densities of the composite sample, resin and magnetite, respectively. For this, the density of the platelike samples was determined with the following procedure:
\begin{enumerate}[$i.$]
\item A photograph of one of the sample's faces was taken along with a reference scale and its area determined by processing the photograph with \emph{ImageJ.}
\item The thickness $\l_\mathrm{s}$ of the sample was measured with a digital micrometer and its volume calculated.
\item Finally, the sample was weighted with an analytical balance and its mass density $\rho$ calculated.
\end{enumerate}
For example, the density measured from our solid magnetite sample is 5.2 g/cm$^3$ that agrees very well with the reported one 5.17 g/cm$^3$ \cite{crc05}. The results obtained for $\eta_\mathrm{m}$, $l_\mathrm{s}$ and $\rho$ are presented in table~\ref{tab:tab1} for each of the samples studied in this work. Given the methods used to measure these quantities, only systematic errors are produced in their determination and thus neglected.

\begin{figure}[t]
\begin{center}
\includegraphics[width=0.5\textwidth]{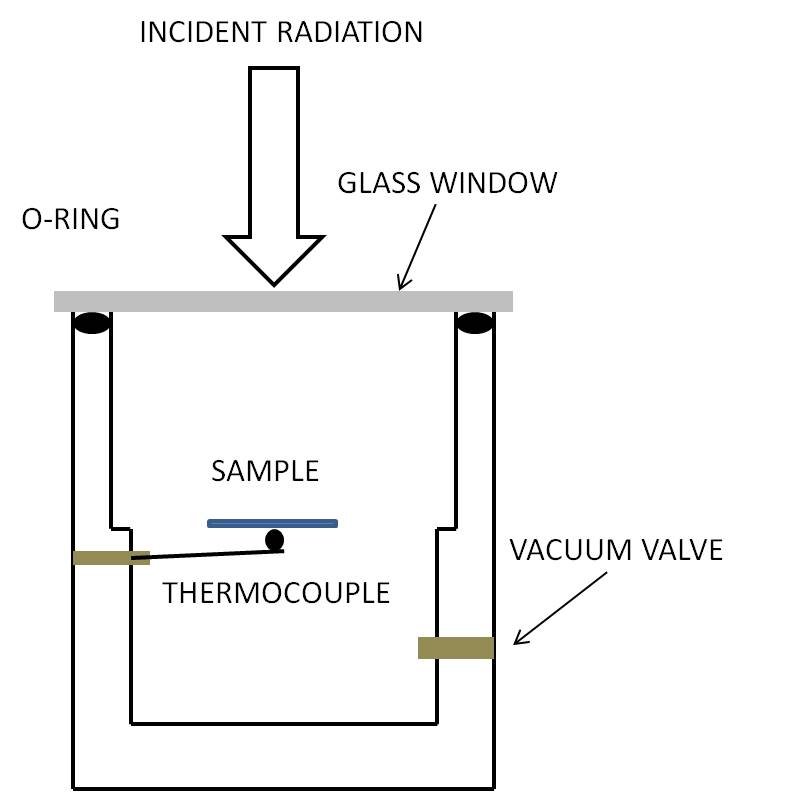} 
\caption{Schematic diagram of the experimental arrangement used to measure the pro\-duct $\rho c$ using the thermal relaxation method (see text for details).}
\label{fig:trset}
\end{center}
\end{figure}

\subsection{Experimental setups}
\label{ssec:exset}

The PA technique used in this work is the well established \emph{open-cell method} widely reported in the literature \cite{dos02, per87}. The thermal diffusivity, $\alpha_\mathrm{s}$, was measured using the experimental setup represented in the schematic diagram of figure~\ref{fig:paset}. In this arrange\-ment, the sample is directly mounted onto a commercial electret microphone (in this case a RadioShack 270-0090). The beam of a 150 W tungsten lamp (Thorlabs OLS1) was focused onto the sample and mechanically modulated with a Stanford Research Systems (SRS) optical chopper model SR540. As a result of the periodic heating of the sample by absorption of the modulated light, the microphone produced a signal that was monitored using a lock-in amplifier (also from SRS model SR530) as a function of the mo\-du\-la\-tion frequency. The illuminated face of the samples was painted with a matte black alkyd enamel in order to ensure a good absorption of light. This coating of paint amounts to about $30 \, \mu \mathrm{m}$ of the thickness of the samples.

The other property we measured is the product $\rho c$ (a.k.a.\ the volumetric heat capacity), corres\-ponding to the product of the mass density and the constant pressure specific heat, respectively. For this, we used the thermal relaxation method \cite{hat79}. Prior to the measurement, both faces of the sample are sprayed with the matte black alkyd enamel in order to make its emissivity approximately equal to one. As shown in the schematic diagram of figure~\ref{fig:trset}, the sample is positioned inside a vacuum chamber --- where partial vacuum has been established --- with one of its faces illuminated by the light beam of a solid-state 100 mW blue laser with a wavelength of \hbox{473 nm}. The temperature of the opposite face (the non-illuminated face) of the sample is traced with Type K bead-wire temperature probe connected to a thermocouple monitor (Extech EA15). As the sample is illuminated, its temperature raises to an equilibrium value above the room temperature. From the behavior of the temperature as a function of time, the product $\rho c$ can be calculated.

\begin{figure}[t]
\begin{center}
\includegraphics[width=0.55\textwidth]{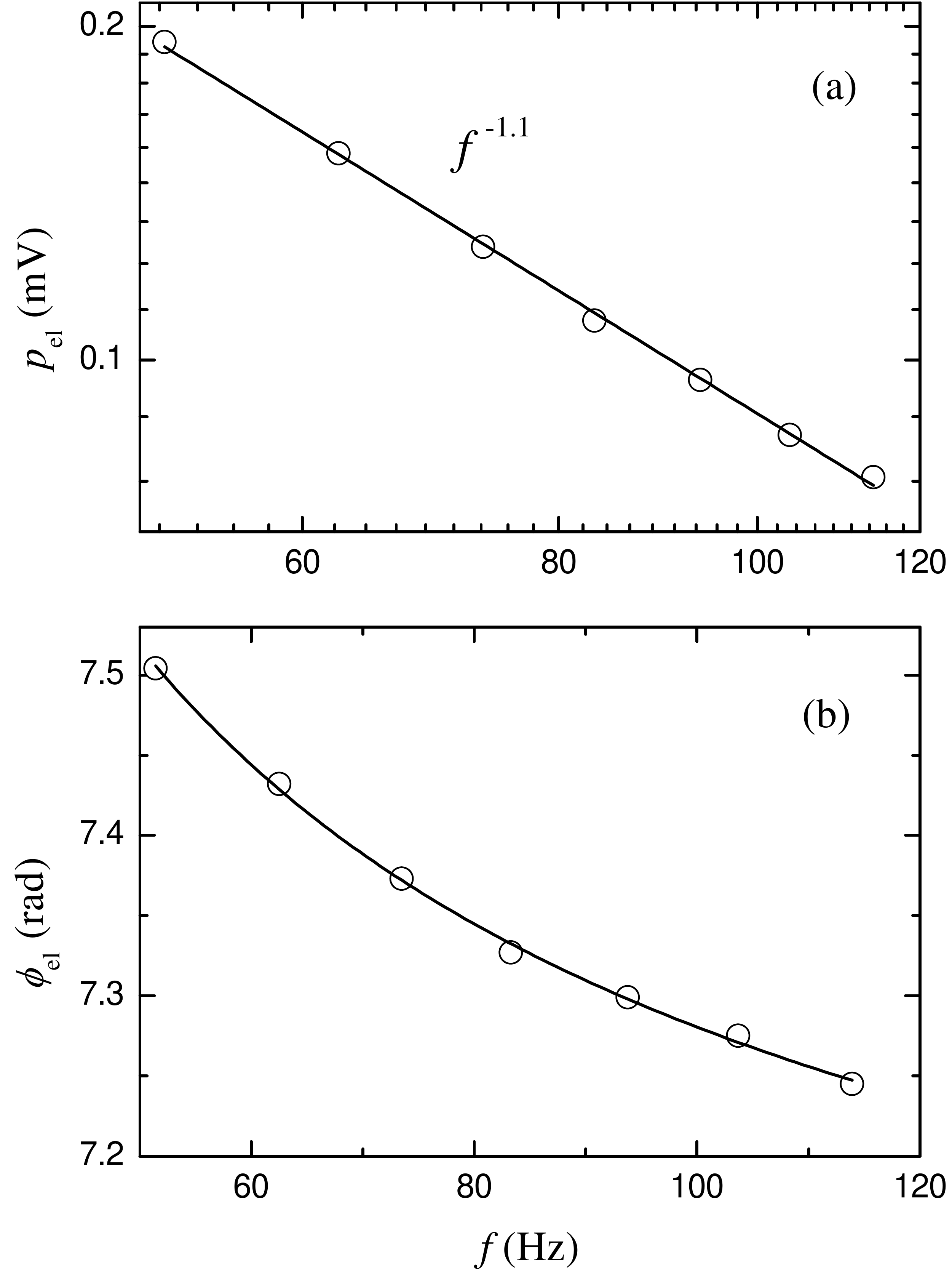}
\caption{Typical dependence of the amplitude $p_\mathrm{el}$ (a) and the phase $\phi_\mathrm{el}$ (b) of the PA signal on the modulation frequency $f$. The clear circles in the log-log plot in (a) correspond to the measured amplitude of the signal in mV, while the solid line corresponds to the fitted slope of the curve. In this case, a slope of $-1.1$ indicates that thermoelastic bending dominates in the generation of the PA signal. The clear circles in (b) correspond to the measured phase of the PA signal while the solid line corresponds to a fit with equation (\ref{eq:phiel}). The results shown here correspond to the isotropic sample with $\eta_\mathrm{m} = 0.076$.}
\label{fig:pa-signal}
\end{center}
\end{figure}

\section{Results and discussion}
\label{sec:resdis}

\subsection{Measurement of the thermal diffusvity}
\label{ssec:mtd}

It is well known that the PA signal, specially that one produced from platelike samples, has two main contributions: one coming from the thermal diffusion phenomenon and the other one from the thermoelastic bending effect \cite{ros76, per87}. For this, there are well established models that allow one to distinguish which one of these contributions dominates. The thermal diffusivity, $\alpha_\mathrm{s}$, is then obtained from the dependence of the detected PA signal on the modu\-lation frequency $f$, as discussed in detail in \cite{dos02, per87}. 

In first place, one must determine if the sample is thermally thin or thick. These two regimes are separated by a cut-off frequency given by $f_\mathrm{c} = \alpha_\mathrm{s}/(\pi l_\mathrm{s}^2)$, where $l_\mathrm{s}$ is the thickness of the sample. A thermally thin sample fulfills the condition $f \ll f_\mathrm{c}$ and the amplitude of the PA signal behaves as $f^{-1.5}$, independent of the properties of the sample. On the other hand, thermally thick samples fulfill the condition $f \gg f_\mathrm{c}$. In this regime, if the thermoelastic bending contribution dominates, the amplitude of the PA signal varies as $p_\mathrm{el} \propto f^{-1}$, while its phase $\phi_\mathrm{el}$ approaches $90^{\circ}$ as
\begin{equation}
\phi_\mathrm{el} \simeq \frac{\pi}{2} + \arctan\left[ \frac{1}{\sqrt{b_\mathrm{s} f} - 1} \right],
\label{eq:phiel}
\end{equation}
where $b_\mathrm{s}$ is the fitting parameter. On the other hand, if the thermal diffusion phenomenon dominates in the generation of the PA signal, the amplitude, $p_\mathrm{td}$, and phase, $\phi_\mathrm{td}$, of the signal have dependences on the modulation frequency of the forms
\begin{equation}
p_\mathrm{td} = \frac{1}{f} \exp\left[ -\sqrt{b_\mathrm{s} f} \right]
\label{eq:diffamp}
\end{equation}
and
\begin{equation}
\phi_\mathrm{td} = -\frac{\pi}{2} - \sqrt{b_\mathrm{s} f}.
\label{eq:diffphase}
\end{equation}
Depending the model used, the thermal diffusivity can be estimated through the relation
\begin{equation}
\alpha_\mathrm{s} = \frac{\pi l_\mathrm{s}^2}{b_\mathrm{s}}.
\label{eq:therm-diff}
\end{equation}

Figure~\ref{fig:pa-signal} shows the typical dependence of the amplitude and phase of the PA signal on the modulation frequency for the composite samples we studied. In particular, the results shown in the figure correspond to an isotropic sample with volume fraction concentration of inclusions $\eta_\mathrm{m} = 0.076$. Notice the slope $-1.1$ of the amplitude $p_\mathrm{el}$ as a function of the modulation frequency $f$ in the log-log plot of figure~\ref{fig:pa-signal}(a). This means that the thermoelastic bending effect dominates in the generation of the PA signal. This was the case for all of our samples, including the pure resin and pure magnetite ones. According to our discussion above, the value of the thermal diffusivity $\alpha_\mathrm{s}$ is obtained by fitting the experimental data for the phase of the PA signal with equation (\ref{eq:phiel}) as shown in figure~\ref{fig:pa-signal}(b).

On an additional note, in order to verify that our experimental setup was operating properly, calibration measurements of the thermal diffusivity of an iron sample with thickness $l_\mathrm{s}=745 \, \mu\mathrm{m}$ were systematically performed. For this particular sample, the thermal diffusion phenomenon dominates in the generation of the PA signal (not shown) and the thermal diffusivity value is obtained by fitting the experimental data with equations (\ref{eq:diffamp}) and (\ref{eq:diffphase}). We measured a value of $2.39 \pm 0.12 \times 10^{-5} \, \mathrm{m}^2/\mathrm{s}$ that agrees with the reported one $2.3 \times 10^{-5} \, \mathrm{m}^2/\mathrm{s}$ \cite{crc05}.

\begin{figure}[t]
\begin{center}
\includegraphics[width=0.65\textwidth]{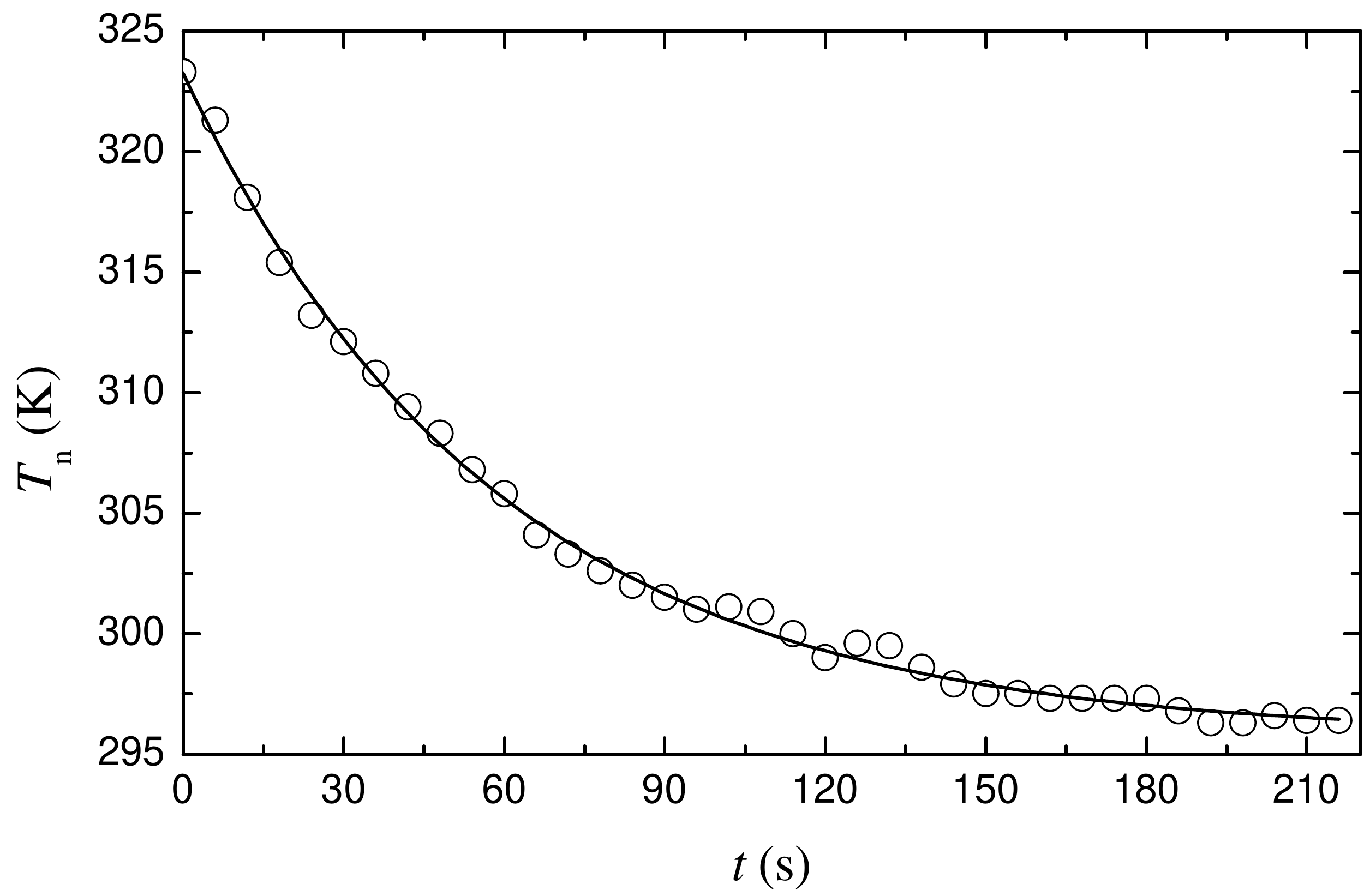} 
\caption{Typical experiment of thermal relaxation for a composite sample. The clear circles correspond to the measured  temperature $T_\mathrm{n}$ of the non-illuminated face of the sample as a function of the time $t$. This results correspond to the isotropic sample with $\eta_\mathrm{m} = 0.076$. The solid line corresponds to a fit with \hbox{equation (\ref{eq:thermdecay})}.}
\label{fig:therm-decay}
\end{center}
\end{figure}

\subsection{Measurement of $\rho c$ and the thermal conductivity}
\label{ssec:mtc}

As shown in the schematic diagram of figure~\ref{fig:trset}, one of the faces of the sample is illuminated  with a constant flux of light, therefore, a lack of equilibrium between the illuminated and non-illuminated faces of the sample is established. This phenomenon can be approximately described by a 1D equation when the thickness $l_\mathrm{s}$ of the sample (including the two coats of black paint) is much smaller than its transverse dimension. The conservation condition for the energy is
\begin{equation}
I_0 - \sigma T_\mathrm{i}^4 - \sigma T_\mathrm{n}^4 = \frac{\mathrm{d}}{\mathrm{d}t} \int_0^{l_\mathrm{s}} \rho c T(x,t) \, \mathrm{d}x,
\label{eq:engcon1}
\end{equation}
where $I_0$ is the flux of incident light over the front face, $\sigma$ is the Stefan-Boltzmann constant, $T_\mathrm{i}$ is the temperature of the \emph{illuminated} face, $T_\mathrm{n}$ is the temperature of the \emph{non-illuminated} (oposite) face, $\rho$ is the mass density of the sample and $c$ is its specific heat at constant pressure. In this equation we use explicitly the fact that the sample is painted with a thin coat of black paint that has an emissivity coefficient approximately equal to one \cite{hat79}.

For long times, when the equilibrium is reached and the fluxes of incident and emitted radiation cancel each other out, the illuminated and non-illuminated faces of the sample reach a saturation temperature $T_\mathrm{i,\,max}$ and $T_\mathrm{n,\,max}$, respectively. Moreover, for the values of $l_\mathrm{s}$ and $I_0$ that we used in the laboratory, the condition $l_\mathrm{s} \, \mathrm{d}T(x,t)/\mathrm{d}x \ll T_\mathrm{i}(t) \approx T_\mathrm{n}(t)$ is fulfilled, thus $T_\mathrm{i,\,max} \approx T_\mathrm{n,\,max}$.

Using the fact that $c$ does not depend on the position and that it is practically constant in the
interval of a few degrees above room temperature, equation (\ref{eq:engcon1}) can be approximately solved for the decrease of temperature of the non-illuminated face of the sample, from $T_\mathrm{n,\,max}$ to $T_\mathrm{n,0}$, after the illumination is interrupted with the resulting expression,
\begin{equation}
T_\mathrm{n}(t) = T_\mathrm{n,0} + (T_\mathrm{n,\,max} - T_\mathrm{n,0}) \exp(-t/\tau_\mathrm{d}).
\label{eq:thermdecay}
\end{equation}
Here, $T_\mathrm{n,0}$ is the initial temperature of the sample before the illumination process starts. It is also the temperature that the sample reaches, for long times, after the illumination is interrupted. In this case, the mean relaxation time $\tau_\mathrm{d}$ is given by
\begin{equation}
\tau_\mathrm{d} = \frac{\rho c l_\mathrm{s}}{8 \sigma T_\mathrm{n,0}^3}.
\label{eq:tau}
\end{equation}
Details of this solution's derivation can be found in \cite{dos02}.

Figure~\ref{fig:therm-decay} shows a typical experiment of thermal relaxation for the composite samples studied here. The results shown in the figure correspond to the measured  temperature $T_\mathrm{n}$ (clear circles) --- as a function of time --- of the non-illuminated face of the isotropic sample with $\eta_\mathrm{m} = 0.076$, while the solid line corresponds to a fit with equation (\ref{eq:thermdecay}). Then, the value of the product $\rho c$ is obtained from equation (\ref{eq:tau}), while the thermal conductivity $k$ is calculated from the very well known relationship
\begin{equation}
k = \rho c \alpha_\mathrm{s}.
\label{eq:thermcond}
\end{equation}

A summary of the results obtained for all of the samples studied in this work is presented in table~\ref{tab:tab1}. The error of the thermal conductivity $k$ was determined by propagating the errors that result  from fitting the corresponding models to the experimental data in the measurements of the thermal diffusivity $\alpha_s$ and the volumetric heat capacity $\rho c$.

\begin{table*}[t]
\centering
\caption{\label{tab:tab1} Summary of the results for the properties of the samples studied in this work. The second block corresponds to the isotropic samples, while the third and fourth to the anisotropic longitudinal and transversal samples, respectively.}
 
\begin{indented}
\lineup
\item[]\begin{tabular}{@{}*{6}{l}}
\br
$\eta_\mathrm{m}$ (v.f.) & $l_\mathrm{s}$ $(\mu\mathrm{m})$ & $\rho$ (kg/m$^3$) & $\alpha_\mathrm{s} \times 10^{-6}$ $(\mathrm{m}^2/\mathrm{s})$ & $\rho c \times 10^{5}$ (J/m$^3$K) & $k$ (W/mK) \cr
\mr
Resin & 1032 & 1100 & 24.65 $\pm$ 1.08 & \08.36 $\pm$ 0.48 & 20.60 $\pm$ 1.48 \cr
Magnetite & \0908 & 5200 & 48.49 $\pm$ 3.60 & 13.30 $\pm$ 0.35 & 64.49 $\pm$ 5.12 \cr
\mr
0.013 & 1047 & 1150 & 28.96 $\pm$ 0.66 & \06.96 $\pm$ 0.19 & 20.14 $\pm$ 0.70 \cr
0.032 & 1020 & 1230 & 29.45 $\pm$ 0.81 & \07.30 $\pm$ 0.26 & 21.49 $\pm$ 0.96 \cr
0.034 & 1066 & 1240 & 31.78 $\pm$ 0.69 & \06.68 $\pm$ 0.17 & 21.21 $\pm$ 0.71 \cr
0.062 & \0973  & 1360 & 29.25 $\pm$ 0.75 & \07.07 $\pm$ 0.27 & 20.67 $\pm$ 0.94 \cr
0.066 & 1024 & 1370 & 29.60 $\pm$ 1.23 & \06.94 $\pm$ 0.16 & 20.55 $\pm$ 0.97 \cr
0.076 & \0980  & 1410 & 29.66 $\pm$ 0.75 & \07.00 $\pm$ 0.18 & 20.76 $\pm$ 0.74 \cr
0.077 & \0980  & 1420 & 30.97 $\pm$ 0.83 & \07.13 $\pm$ 0.20 & 22.06 $\pm$ 0.85 \cr
0.089 & \0895  & 1470 & 24.85 $\pm$ 0.49 & \09.10 $\pm$ 0.26 & 22.63 $\pm$ 0.78 \cr
\mr
0.016 & \0962  & 1168 & 30.90 $\pm$ 0.62 & \06.84 $\pm$ 0.22 & 21.13 $\pm$ 0.81 \cr
0.042 & 1014 & 1273 & 31.66 $\pm$ 1.52 & \06.73 $\pm$ 0.22 & 21.31 $\pm$ 0.81 \cr
0.049 & \0996  & 1304 & 31.86 $\pm$ 0.44 & \06.73 $\pm$ 0.19 & 21.44 $\pm$ 0.66 \cr
0.061 & \0953 & 1352 & 28.25 $\pm$ 0.99 & \08.14 $\pm$ 0.14 & 22.98 $\pm$ 0.89 \cr
0.064 & \0956  & 1364 & 22.43 $\pm$ 0.29 & \06.99 $\pm$ 0.15 & 15.67 $\pm$ 0.40 \cr
0.069 & \0998  & 1383 & 22.91 $\pm$ 0.51 & \06.33 $\pm$ 0.18 & 14.49 $\pm$ 0.51 \cr
\mr
0.014 & \0805  & 1158 & 22.62 $\pm$ 0.31 & \07.20 $\pm$ 0.20 & 16.28 $\pm$ 0.50 \cr
0.023 & \0960  & 1194 & 31.60 $\pm$ 1.52 & \05.47 $\pm$ 0.19 & 17.27 $\pm$ 1.02 \cr
0.033 & \0982  & 1238 & 33.51 $\pm$ 0.42 & \05.77 $\pm$ 0.18 & 19.34 $\pm$ 0.66 \cr
0.039 & \0900  & 1261 & 30.29 $\pm$ 0.28 & \06.79 $\pm$ 0.23 & 20.57 $\pm$ 0.73 \cr
0.041 & \0981  & 1271 & 28.41 $\pm$ 0.69 & \06.58 $\pm$ 0.16 & 18.69 $\pm$ 0.64 \cr
0.044 & 1002 & 1281 & 23.73 $\pm$ 0.56 & \07.38 $\pm$ 0.19 & 17.50 $\pm$ 0.60 \cr
0.051 & \0995  & 1310 & 19.08 $\pm$ 0.98 & \06.56 $\pm$ 0.18 & 12.51 $\pm$ 0.72 \cr
0.052 & \0995  & 1316 & 15.95 $\pm$ 0.69 & \06.94 $\pm$ 0.23 & 11.06 $\pm$ 0.60 \cr
\br
\end{tabular}
\end{indented}
\end{table*}

\begin{figure}[tb]
\begin{center}
\includegraphics[width=0.6\textwidth]{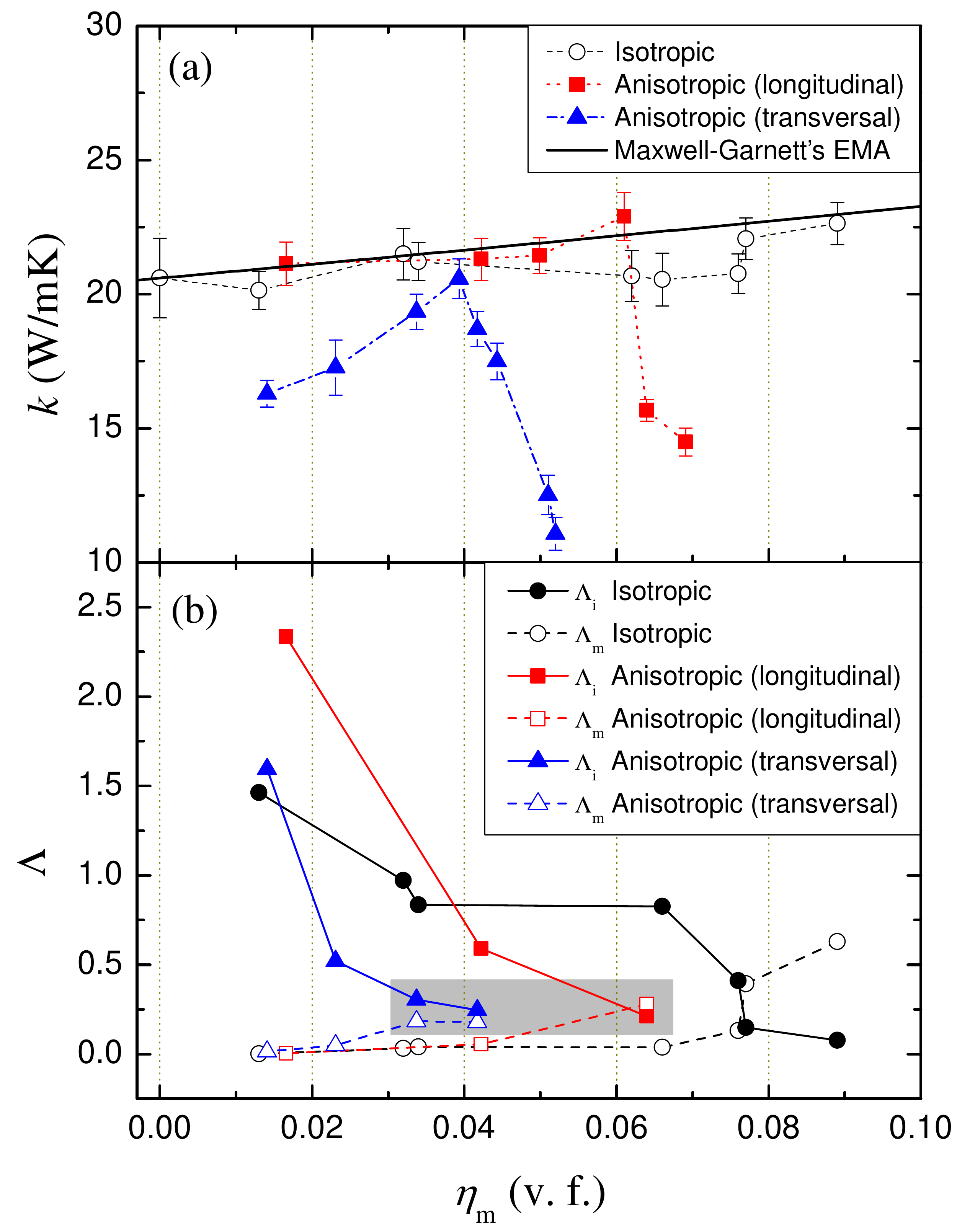}
\caption{In (a), the curves with broken lines correspond to the measured thermal conductivity $k$ of the samples studied here as a function $\eta_\mathrm{m}$. The transversal bars correspond to the propagated experimental error, while the black solid curve corresponds to the Maxwell-Garnett's EMA given in equations (\ref{ema1}) and (\ref{ema2}). Notice the good correspondence with the isotropic samples and the non-trivial dependance of $k$ on $\eta_\mathrm{m}$ for the anisotropic samples. (b) Mean lacunarity $\Lambda$ for selected samples. The solid curves with solid symbols corres\-pond to the mean lacunarity of the inclusion structure, $\Lambda_\mathrm{i}$, while the dashed curves with clear symbols to the mean lacunarity of the matrix, $\Lambda_\mathrm{m}$. Notice how $\Lambda_\mathrm{i}$ or $\Lambda_\mathrm{m}$ remains very low, while the other one is high for isotropic samples. In contrast, for anisotropic (longitudinal and transversal) samples with large enough concentrations (those inside the grey square in the figure), that cannot be described by the standard Maxwell-Garnett's EMA, $\Lambda_\mathrm{i}$ and $\Lambda_\mathrm{m}$ are both moderate and commensurate due to the development of inclusion and matrix domains (see text for details).}
\label{fig:tclac-results}
\end{center}
\end{figure}

\subsection{Effective media approximation}
\label{ssec:ema}

Since its invention, the effective media approximation (EMA) has been employed in the study of several kinds of macroscopically inhomogeneous media. For example, the EMA can be used to estimate the effective properties of systems where a random distribution of inclusions is embedded in a given continuous matrix \cite{str98}. Some of these properties include the dielectric function, elastic modulus, electrical conductivity and the thermal conductivity, being the latter the one we are interested on in this work. If the inhomogeneities inside the medium are large enough, such that in each space portion of the material the behavior of the property is described or controlled by macroscopic constitutive equations, it is only necessary to find a reasonable way of averaging the existing statistical variations in the material. Conversely, the same basic problem arises in different fields such as heat flow, diffusion and elastic properties \cite{lan78, bjo94, mer99, gao06, mat86}.

In this work, we use the Maxwell-Garnett approximation, one of the most commonly used approximation methods. By nature of its construction, this approximation is exact only at first order in volume fraction. At second order, there are big discrepancies with an error up to $50\%$ \cite{cho99}. In the case of composite materials, one evaluates their effective properties as a function of the properties of their components. Here, the Maxwell-Garnett approximation is applied to systems of two components, where one of them is considered as a continuum (polyester resin) that supports the other one (magnetite inclusions). Additionally, our composite samples were prepared in a regime of low concentration. In order to evaluate their effective thermal conductivity we used the following equation:

\begin{equation}
k = k_\mathrm{r} + \frac{3 \eta_\mathrm{m} \gamma}{1 - \eta_\mathrm{m} \gamma} k_\mathrm{r}
\label{ema1}
\end{equation}

\noindent
with
\begin{equation}
\gamma = \frac{k_\mathrm{m} - k_\mathrm{r}}{k_\mathrm{m} + 2 k_\mathrm{r}},
\label{ema2}
\end{equation}
where $k_\mathrm{m}$ and $k_\mathrm{r}$ are the magnetite and polyester resin thermal conductivities, respectively, while $\eta_\mathrm{m}$ is the volume fraction occupied by the magnetite particles \cite{cho99}. In particular, these expressions should be able to predict the effective thermal conductivity of the isotropic samples where the inclusions are randomly distributed inside the matrix.

The broken curves with symbols in figure~\ref{fig:tclac-results}(a) show the measured thermal conductivity $k$ for the isotropic (black dashed line with clear circles) samples and for the anisotropic longitudinal (red dotted line with solid squares) and transversal (blue dash-dotted line with solid triangles) samples as a function of the volume fraction concentration of magnetite particles, $\eta_\mathrm{m}$. The black solid line corresponds to the Maxwell-Garnett's EMA given in equations (\ref{ema1}) and (\ref{ema2}). As expected, our results show a good agreement between the experimental measurements for the isotropic samples with this approximation. Given the fact that magnetite has a larger (more than two times according to table~\ref{tab:tab1}) thermal conductivity than the polyester resin matrix, the effective thermal conductivity of these samples should increase with the concentration of inclusions. Recent results on the development of numerical tools for mesoscopic systems have even extended the range of prediction for effective media theories to include composite materials where the concentration of inclusions is not necessarily low, as long as the inclusions remain randomly distributed. In these conditions, the effective thermal conductivity of composite samples still increases with the concentration of inclusions \cite{wan07}.

Contrastingly, the anisotropic longitudinal samples show a drop in the thermal conductivity for $\eta_\mathrm{m}$ above $0.061$, while the anisotropic transversal samples show a non-trivial behavior with $\eta_\mathrm{m}$ as shown in figure~\ref{fig:tclac-results}(a) with lower thermal conductivities than the isotropic ones. It has been observed that disorder on crystallographic sites largely controls the thermal conductivity of mixed crystals and that mass disorder increases the anharmonic scattering of phonons through a modification of their vibration eigenmodes, resulting in an increase of the thermal resistivity \cite{gie02, alb01, gar11}. Moreover, a model for anisotropic periodic composites with wire and cylindrical inclusions (nanowires and cylindrical nanowires), where the heat flow is applied in the longitudinal direction of the inclusions, predicts a decrease in the effective thermal conductivity due to the introduction of additional surface scattering in relation to the ballistic phonon transport \cite{yan05}. Even though our composite samples are not crystalline, the formation of large \emph{quasi-continuous} structures of inclusions and the development of \emph{domains} in some of the anisotropic samples may explain the observed decrease in their effective thermal conductivity. In particular, the transversal section of the magnetite chains in anisotropic longitudinal samples increases with $\eta_\mathrm{m}$ as can be appreciated in figures~\ref{fig:ImageJ}(e) and \ref{fig:ImageJ}(f). Nonetheless, in samples with $\eta_\mathrm{m} \le 0.061$, these structures are isolated and commensurate with those present in isotropic samples with similar concentrations. Moreover, the magnetite chains are more evenly distributed inside the matrix than in the isotropic samples themselves. For these samples, a good correspondence with the EMA is observed --- for some samples it is even better than that of similar isotropic samples! In contrast, it is only in the anisotropic longitudinal samples with concentration $\eta_\mathrm{m} > 0.061$ where the inclusion structure has a counterintuitive effect, with a sharp drop in their thermal conductivity. As apparent in figure~\ref{fig:ImageJ}(f), above this ``critical'' concentration, the transversal section of the magnetite chains becomes so large that the chains are able to aggregate laterally with the consequent formation of elongated structures and the development of domains of magnetite and resin. Obviously, this kind of samples cannot be described with the Maxwell-Garnett's EMA as the inclusions are not randomly distributed anymore. For these, an analysis of the inclusion structure is provided below.

\subsection{Inclusion structure and thermal conduction}
\label{ssec:inc-struct}

In order to explain the decrease in thermal conductivity for some of the anisotropic samples, we analyzed the multifractality and lacunarity of some of them.

\begin{figure}[tb]
\begin{center}
\includegraphics[width=0.65\textwidth]{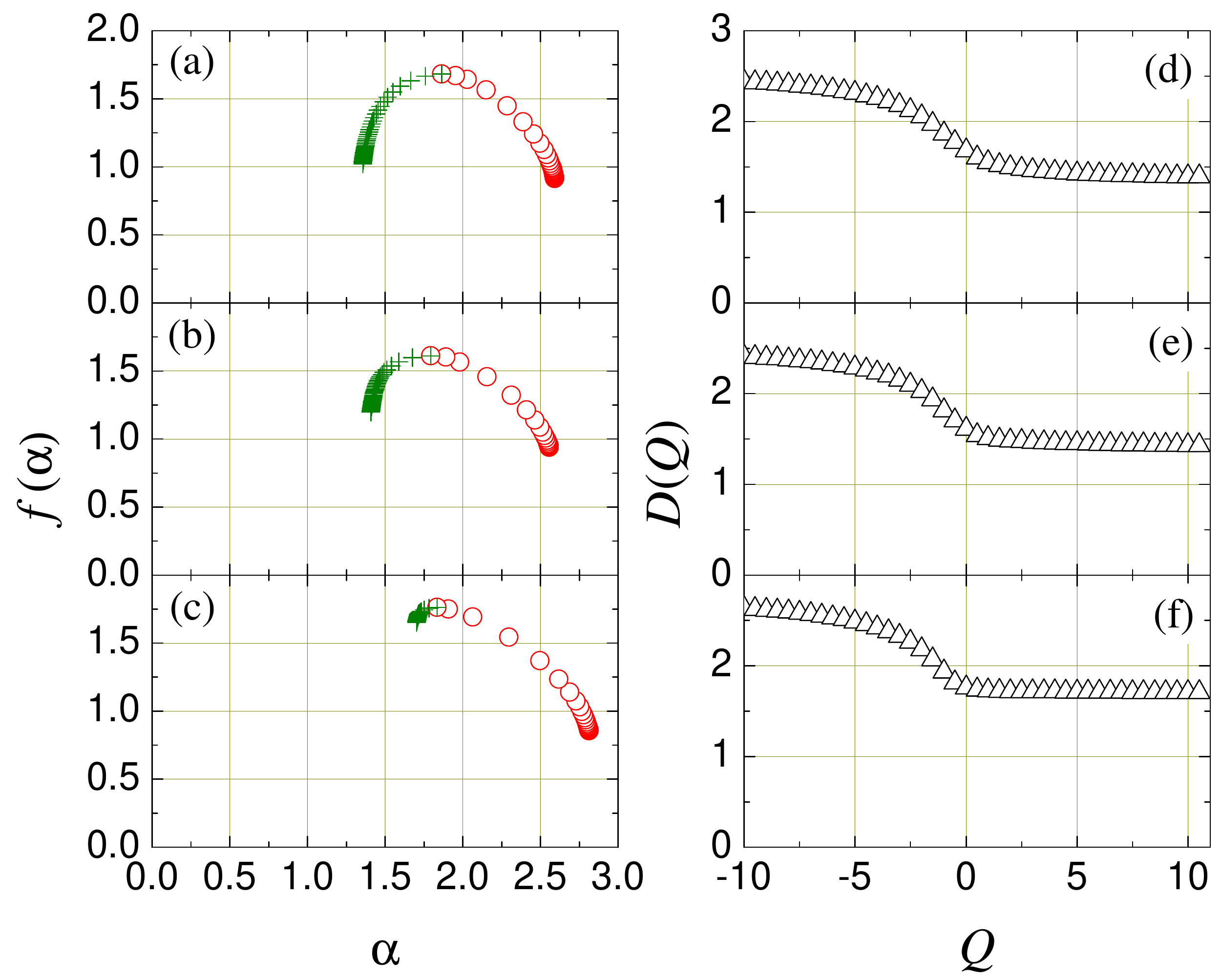}
\caption{Typical singularity spectrum $f(\alpha)$ and generalized fractal dimension $D(Q)$ for an isotropic sample, (a) and (d), an anisotropic longitudinal sample, (b) and (e), and an anisotropic transversal sample, (c) and (f), respectively. Notice the humped shape of $f(\alpha)$ for the isotropic and the anisotropic longitudinal samples in (a) and (b). This fact in combination with the slight negative slope of $D(Q)$ for $Q>0$ in (d) and (e), are signatures of a multifractal inclusion structure. In contrast, anisotropic transversal samples show a convergent and truncated $f(\alpha)$, and a flatter dependence of $D(Q)$ on $Q$ for $Q>0$ shown in (c) and (f), respectively. These characteristics are a signature of a monofractal structure of inclusion. In particular, the results shown here correspond to the isotropic sample with $\eta_\mathrm{m} = 0.034$ in (a) and (d), the anisotropic longitudinal sample with $\eta_\mathrm{m} = 0.042$ in (b) and (e), and the anisotropic transversal sample with $\eta_\mathrm{m} = 0.033$ in (c) and (f).}
\label{fig:mfrac}
\end{center}
\end{figure}

The typical singularity spectrum, $f(\alpha)$, and generalized fractal dimension, $D(Q)$, are presented for isotropic samples in figures~\ref{fig:mfrac}(a) and \ref{fig:mfrac}(d), for anisotropic longitudinal samples in figures~\ref{fig:mfrac}(b) and \ref{fig:mfrac}(e), and for anisotropic transversal samples in figures~\ref{fig:mfrac}(c) and \ref{fig:mfrac}(f), respectively. These results were obtained using the plugin \emph{FracLac} for \emph{ImageJ} from binarized images such as those presented in the even rows of figure~\ref{fig:ImageJ}. In these images, the black patterns correspond to the inclusion structure while the white areas to the resin matrix. This open software calculates both quantities using the box counting method, that has been successfully applied for this kind of analysis in other systems \cite{chh89, pos01}. When the image analyzed has a pattern with multifractal properties, the function $D(Q)$ vs.\ $Q$ is decreasing and sigmoidal around $Q=0$. In our samples, this occurs for the isotropic and anisotropic longitudinal ones, exemplified in figures~\ref{fig:mfrac}(d) and \ref{fig:mfrac}(e). In the case of anisotropic transversal samples, they show a monofractal pattern as the plot for $D(Q)$ tends to be horizontal for $Q>0$ as shown in figure~\ref{fig:mfrac}(f).

The generalized fractal dimension is usually used in combination with other multifractal measures such as the singularity spectrum $f(\alpha)$. The latter is used to characterize the variety within a pattern regarding the scale at which the pattern is observed. Monofractals show less variation than multifractals in their dependence of $f(\alpha)$ on $\alpha$. The singularity spectrum for a monofractal (see figure~\ref{fig:mfrac}(c)) converges to a value while the spectrum for a multifractal is typically humped (see figures~\ref{fig:mfrac}(a) and \ref{fig:mfrac}(b)). From the behavior of the singularity spectra obtained for our samples, the results obtained from the generalized fractal dimension are confirmed in that the inclusion structure of isotropic and anisotropic longitudinal samples have multifractal characteristics, while this structure in anisotropic transversal samples is monofractal.

Additionally, the mean lacunarity of the inclusion structure, $\Lambda_\mathrm{i}$, and that of the matrix, $\Lambda_\mathrm{m}$, were obtained with \emph{FracLac} from the binarized images. The results are shown in figure~\ref{fig:tclac-results}(b). In short, a lacunarity analysis is a multiscaled method for describing patterns of spatial dispersion \cite{plo96}. In contrast with measures such as the fractal dimension, that describes how much space is filled, lacunarity indicates how the space is filled. In this sense, lacunarity is a parameter that describes the distribution of the sizes of gaps or \emph{lacunae} in a give structure. Greater lacunarity reflects a greater size distribution of the lacunae or, said in another way, a higher degree of ``gappiness'' although it has been also defined as visual texture, inhomogeneity, translational and rotational invariance, etc. Nonetheless, lacunarity pertains to both gaps and heterogeneity, and has been applied to the study of many different systems including microvascular ones \cite{gou11}.

In figure~\ref{fig:tclac-results}(b) one can appreciate that, for isotropic samples, $\Lambda_\mathrm{i}$ (black solid curve with solid circles) or $\Lambda_\mathrm{m}$ (black dashed curve with clear cirles) remains very low while the other one is high. This is also the case for the anisotropic longitudinal samples with thermal conductivities in agreement with the EMA --- for these, $\eta_\mathrm{m} \le 0.061$ --- where $\Lambda_\mathrm{i}$ (red solid curve with solid squares) is large while $\Lambda_\mathrm{m}$ (red dashed curve with clear squares) is very low. We associate this behavior with the fact that the inclusions or inclusion structures formed inside the matrix are randomly distributed, thus fulfilling the conditions of the EMA used in this work. In consequence, either the inclusion structure or the matrix retains a high degree of homogeneity and translational and rotational invariance. On the other hand, even when some of the anisotropic transversal samples with the smaller $\eta_\mathrm{m}$ show this behavior, where $\Lambda_\mathrm{i}$ (blue solid curve with solid triangles) is large and $\Lambda_\mathrm{m}$ (blue dashed curve with clear triangles) very small, these samples along with the rest of the anisotropic transversal ones cannot be described by the EMA, as they do not fulfill the condition of having randomly distributed inclusions. This is obviously due to the formation of large chains of magnetite particles across the matrix, all aligned in the same direction.

It is interesting to notice that anisotropic longitudinal samples with $\eta_\mathrm{m} > 0.061$, have moderate values for $\Lambda_\mathrm{i}$ and $\Lambda_\mathrm{m}$ that are commensurate. This is also the case for anisotropic transversal samples as $\eta_\mathrm{m}$ increases, behaviour that is pointed out in figure~\ref{fig:tclac-results}(b) for the data inside the grey square, that can be associated with the development of domains of inclusions and matrix. See, for example, figures~\ref{fig:ImageJ}(f), \ref{fig:ImageJ}(g) and \ref{fig:ImageJ}(h) where one can identify large and elongated structures of magnetite particles intertwined with portions of matrix with zero or very low concentration of inclusions. From the point of view of the lacunarity analysis, this structure retains a moderate degree of homogeneity and translational and rotational invariance both, in the inclusion structure and in the matrix itself. At this point, it is necessary to remind us that the samples from which these images were obtained are very thin ($\sim 100 \, \mu\mathrm{m}$) and only depict a few layers of inclusion structure. One must extrapolate this fact to the thicker \hbox{($\sim$ 1 mm)} samples that were thermally characterized, where domains of inclusion structures and matrix overlap disorderly, each one presenting a high thermal impedance or resistivity to the one below, thus quenching the thermal transport in the direction of illumination. Our observation is consistent with the prediction of the model developed for fractal-like tree networks proposed by Yu and Li \cite{yu06}. They found that the effective thermal conductivity of composites with this kind of inclusion networks, decreases with an increase of the length of the branches or density of the network. Moreover, they conclude that thermal conductivity of the network itself may be less than that of the original material by several orders of magnitude. For our composite samples that present long chains of inclusions in a fractal or multifractal structure this seems to be the case, with a drop in their effective thermal conductivity, as it is apparent that the inclusion structure has a lower thermal conductivity than that of the magnetite and resin themselves.

\section{Conclusions}
\label{conc}

In this paper, we have studied the thermal conductivity of isotropic and anisotropic samples consisting in a polyester resin matrix with embedded inclusions of magnetite particles in a fractal structure. Our results show that all of the isotropic samples and some of the anisotropic longitudinal ones, where the distribution of inclusions or inclusion structures (with these structures isolated from each other) remains random, present an effective thermal conductivity that can be described by the standard Maxwell-Garnett effective medium approximation. In contrast, some of the anisotropic longitudinal samples and all of the anisotropic transversal ones, where the development of long chains of magnetite particles leads to the formation of magnetite and resin domains in a multilayered fashion, show a drop in their measured thermal con\-duc\-ti\-vi\-ty with values lower than that of the isotropic ones. These samples cannot be described by the Maxwell-Garnett's approximation, however, our results are consistent with the work of Yu and Li \cite{yu06} on the effective thermal conductivity of composite samples with a fractal inclusion structure. Our study indicates that the thermal conductivity of the magnetite inclusion structure has a lower thermal conductivity than that of the magnetite and resin themselves, making the effective thermal conductivity of these samples highly dependent on the structural properties of the inclusion arrangement. Results of this study can be used to direct the development of composite materials with tuned effective thermal conductivities by controlling the aggregation processes in the inclusion structure.

\ack

This work was partially supported by CONACyT and by SEP through the grant PROMEP/103.5/10/7296. The authors are thankful to R. Silva-Gonz\'alez (BUAP) for the SEM micrographs and to M. E. Mendoza (BUAP) for the XRD measurements. VD is thankful to V. M. Kenkre (UNM) for his hospitality.

\end{document}